\documentclass[usenatbib]{mn2e}   % MNRAS

\usepackage{natbib} 
\usepackage{graphicx}
\usepackage{times}                % MNRAS
 
\newcommand{\aap}{A\&A}           % MNRAS

\newcommand{\apj}{ApJ}

\newcommand{\mnras}{MNRAS}
\newcommand{\ion}[2]{{#1}{\sc{#2}}}
\newcommand{\object}[1]{{#1}}
\newcommand{\FBDS}{{\tt A$^\mathtt{sf}$}}
\newcommand{\FBDV}{{\tt A$^\mathtt{sf}_{400}$}}
\newcommand{\FBDW}{{\tt A$^\mathtt{sf}_{600}$}}
\newcommand{\FBDX}{{\tt A$^\mathtt{sf}_{1000}$}}
\newcommand{\FBDT}{{\tt A$^\mathtt{nosf}$}}
\newcommand{\EACB}{{\tt H}}

\newcommand{\sdrop}{$\sigma$-drop}
\newcommand{\sdrops}{$\sigma$-drops}
\newcommand{\sold}{$\sigma^\mathrm{old}_R$}
\newcommand{\snew}{$\sigma^\mathrm{new}_R$}
\newcommand{\smin}{$\sigma^\mathrm{min}_R$}
\newcommand{\smax}{$\sigma^\mathrm{max}_R$}
\newcommand{\sfrunit}{M$_{\sun}\ \mbox{yr}^{-1}$}
\newcommand{\kms}{km$\,$s$^{-1}$}
\newcommand{\slos}{$\sigma_\mathrm{los}$}
\newcommand{\CO}{\element[][12]{CO}\ }   % A&A

\renewcommand{\CO}{{$^{12}$CO\ }}

\title{Lifetime of nuclear velocity dispersion drops in barred galaxies}
\author[H. Wozniak and N. Champavert]
{Herv\'e Wozniak and Nicolas Champavert\\
Centre de Recherche Astronomique de Lyon, 9 avenue Charles Andr\'e, F-69561 Saint-Genis Laval cedex, France
}

\begin{document}

% Version MNRAS
\date{Accepted 2005 XXX XXX. Received 2005 XX XX ; in original form 2005 08 23}
%\pagerange{\pageref{firstpage}--\pageref{lastpage}} \pubyear{2005}
\maketitle
%\label{firstpage}  

\begin{abstract}  % MNRAS
We have made hydro/N-body simulations with and without star formation
to shed some light on the conditions under which a central
kinematically cold stellar component (characterized by a velocity
dispersion drop or \sdrop) could be created in a hot medium (e.g. a
bulge) and survive enough time to be observed.  We found that the
timescale for a \sdrop\ formation could be short (less than 500~Myr)
whereas its lifetime could be long (more than 1~Gyr) provided that the
central region is continuously or regularly fed by fresh gas which
leads to a continuous star formation activity. Star formation in the
central region, even at a low rate as 1~\sfrunit, is mandatory to
sustain a permanent \sdrop\ by replacing heated particles by new
low-$\sigma$ ones.  We moreover show that as soon as star formation is
switched off, the \sdrop\ begins to disappear.

% Version MNRAS
\end{abstract}

%--------------------------------------------------------------------
% INTRODUCTION
%--------------------------------------------------------------------
\section{Introduction}

The discovery of a sub-kpc central drop in stellar velocity dispersion
(\sdrop\ in short) in three Seyfert galaxies with double bars
(\citealt{Eetal01}, hereafter Paper~I) has brought this phenomenon to
the foreground. In fact, a \sdrop\ was previously observed in a single
object (\object{NGC\,6503}) by \citet{Bott89}. To explain the low
velocity dispersion in the centre, \citet{Bott97} had to built up a
model with a distinct dynamically cold and compact isothermal nucleus
in addition to the large scale stellar disc with the same
mass-to-light ratio. Others examples can be found in a few papers
(\citealt{HS98}; \citealt{HSMP99}; \citealt{JDMB88} for instance) but
have been most probably overlooked because of the low signal-to-noise
ratio of these observations. More recent observations of larger sample
and better signal-to-noise ratio have confirmed (\citealt{Metal03},
\citealt{n1068}, Emsellem et al. in preparation) the presence of
\sdrop\ in several single-barred galaxies. However, it is still
premature to attempt to make statistics on the frequency of \sdrops\
in galaxies since a complete and systematic survey is not yet
available. More challenging and not yet explained is the observation
of \sdrops\ in elliptical galaxies (e.g. \object{NGC\,1404},
\citealt{GCBZL98}). The occurrence of \sdrops\ in such objects should
deserve a detailed study.

\citet{Wetal03}(hereafter Paper~II) found that the
stellar \slos\ drop is the kinematical signature of stars born from a
dynamically cold gaseous component. The origin of the drop results
from the conspiracy of two factors.  First, the new stars are
concentrated toward the centre, and since they are just born from the
low-dispersion gas component, their dispersion is much lower than for
the old stars component. But second, the effect is amplified by the
fact that the gas dispersion is also dropping toward the centre (and
therefore the new stellar component too). This is due to the strong
accumulation of gas toward the centre, in a nuclear disk, where
dissipation is stronger than elsewhere, and where the gas is therefore
cooling down efficiently.

%v3.0
The existence of flattened and rapidly rotating nuclear stellar
discs has been predicted by \citet{SB89} and further studied by
\citet{SBF90}.  They showed that such stellar discs could remain
flattened for a long time since two body relaxation is a slow process.

Thus, a few issues remain opened:
\begin{enumerate}

\item The frequency of \sdrop\ in galaxies depends on the lifetime of
the phenomenon. In Paper~II we were not able to give such an
estimate. However, even if the frequency of this kinematical effect is
still under unknown, it can be firmly stated that it is not a peculiar
phenomenon since it is observed in an increasing number of `normal'
galaxies. We give in Table~\ref{tab:census} an updated list of \sdrop\
observations. It is worth noting that \sdrops\ have been detected
using a number of absorption lines from the optical to the near
infrared (Mg line triplet at $\lambda\lambda$ 5164, 5173, 5184 \AA,
\ion{Ca}{ii} triplet at $\lambda$ 8542 and 8662 \AA, \CO\ molecular
bands at roughly 2.29$\mu$m, etc.) which traces various ages of
stellar populations. Moreover, \sdrops\ cannot be data reduction
artifacts since stellar kinematics has been retrieved from absorption
lines using various techniques.

\item Other mechanisms could be at work to remove some kinetic energy
from the stellar component. For instance, a massive and concentrated
dark halo seems to be able to generate a \sdrop\ in one of the
simulations of \citet{AM02} (their Fig.~13).

\item Beyond its intrinsic dynamical lifetime, the detectability of
any \sdrop\ should be also limited by the lifetime of the stellar
populations that have been used to trace the kinematics. For instance,
in Paper~I, \sdrops\ has been detected using the $^{12}$CO bandhead
at 2.29~$\mu$m.

\end{enumerate}

In this paper, we will specifically address the problem of \sdrop\ lifetime using a
new set of N-body experiments.

\begin{table*}
\caption{List of confirmed or suspect \sdrops\ in the literature.}
\label{tab:census}
\centering
%\begin{minipage}{140mm}
\begin{tabular}{@{}llllllll@{}}
\hline
%\hline
Object     &    Morph. Type  & Nucleus & Reference                            &Main abs. lines & Technique \\
\hline
NGC 1068   &	SA(rs)b	   &   Sy 1/2  &	Emsellem et al. 2005          & Mg b          & pPXF$^1$\\
NGC 1097   &	SB(s)b	   &    Sy 1   &	Emsellem et al. 2001          & \CO\ 2.3$\mu$m& pPXF \\
NGC 1138   &	SBO	   &	       &	Simien \& Prugniel 2002       & Mg b          & FFM$^2$\\
NGC 1808   &	SAB(s:)b   &	Sy 2   &	Emsellem et al. 2001          & \CO\ 2.3$\mu$m& pPXF \\
NGC 2639   &	SA(r)a	   &    Sy 1.9 &	Marquez et al. 2003           & CaII T        & CCF\\
NGC 3021   &    SAbc       &           &        H\'eraudeau et al. 2003       & Mg b          & FFM \\
NGC 3412   &	SB0(s)	   &	       &	Aguerri et al. 2003           & Mg b          & FCQ$^3$ \\
NGC 3623   &	SAB(rs)a   &	Liner  &	De Zeeuw et al. 2002          & Mg b          & FCQ \\
NGC 3627   &	SBb        &Liner/Sy 2 &	H\'eraudeau \& Simien 1998    & Mg b          & FFM \\
NGC 4303   &	SBbc       &	Sy 2   &	H\'eraudeau \& Simien 1998    & Mg b          & FFM \\
NGC 4579   &	SBb        &Liner/Sy 1.9&	H\'eraudeau \& Simien 1998    & Mg b          & FFM \\
NGC 4594   &    SA(s)a     &    Sy 1.9 &        Jarvis \& Dubath 1988         & Mg b          & CCF$^4$ \\
NGC 4725   &	SBab       &	Sy 2   &	H\'eraudeau et al. 1999       & Mg b          & FFM \\
NGC 4477   &	SB0        &	Sy 2   &	Jarvis et al. 1988            & Fe, Mg b      & CCF \\
NGC 5728   &	SABb(r)    &	Sy 2   &	Emsellem et al. 2001          & \CO\ 2.3$\mu$m& pPXF \\
NGC 6503   &	SA(s)cd	   &Liner/HII  &	Bottema 1989; Bottema \& Gerritsen 1997 & Fe, Mg b & CCF \\
NGC 6814   &	SAB(rs)bc  &	Sy 1.5 &	Marquez et al. 2003           & CaII T        & CCF\\
NGC 6951   &	SAB(rs)bc  &	Sy 2   &	Marquez et al. 2003           & CaII T        & CCF\\
NGC 7177   &    SBb        &Liner/HII  &        H\'eraudeau et al. 2003       & Mg b          & FFM \\
IC 184     &	SB(r)a	   &    Sy 2   &	Marquez et al. 2003           & CaII T        & CCF\\
Circinus   &    SAb        &    Sy 2   &        Maiolino et al. 1998          & \CO\ 2.3$\mu$m& CCF \\
\hline
NGC 3593$\dag$&	SA(s)0/a   &	Sy 2   &        Bertola et al. 1996           & 5200--6200\AA & FCQ \\
\hline
\end{tabular}

\flushleft
$\dag$ edge-on galaxy\\
$^1$ Penalized pixel fitting \citep{CE04} \\
$^2$ Fourier Fitting Method \citep{FIH89}\\
$^3$ Fourier Correlation Quotient \citep{B90}\\
$^4$ Cross-Correlation Function (e.g. \citealt{Bott88} and reference therein)\\
%\end{minipage}
\end{table*}

%Ca ii absorption triplet at 8542; 8662 \AA ,
% Mg line triple ($\lambda\lambda$ 5164, 5173, 5184 \AA)

%-------------------------------------------------------------------
% DESCRIPTION DE LA TECHNIQUE
%-------------------------------------------------------------------

\section{Description of the numerical experiments}
\label{sec:model}

An initial stellar population is set up to reproduce a typical disc
galaxy. Positions and velocities for $2.5\,10^6$ particles are drawn
from a superposition of two axisymmetric \citet{MN75} discs of mass
respectively $M_1$ and $M_2$ (cf. Table~\ref{tab:simul}), of scale
lengths respectively $1$ and $3.5$~kpc and common scale height of
$0.5$~kpc. Initial velocity dispersions are computed solving
numerically the Jeans equations. 
%For sensible choices of $M_1/M_2$, a
%radial velocity dispersion ($\sigma_R$) drop can be self-consistently
%set up. 
The initial disc radius is 30~kpc except for run \EACB\ for
which the truncation radius is 50~kpc.
The gaseous component of runs \FBDS, \FBDV, \FBDW\ and \FBDX\ is
represented by 50\,000 particles for a total mass of
$1.1\,10^{10}$~M$_{\sun}$ distributed in a 6~kpc scalelength
Miyamoto-Nagai disc. For the collisionless run \FBDT\, $M_1$ and $M_2$
have been proportionally scaled so as to keep the same total mass than
\FBDS.

Evolution is computed with a particle--mesh N-body code which includes
stars, gas and recipes to simulate star formation when necessary. The
broad outlines of the code are the following: the gravitational forces
are computed with a particle--mesh method using a 3D polar grid with
$(N_R, N_\phi, N_Z)=(31,32,64)$ active cells, the hydrodynamics
equations are solved using the SPH technique and the star formation
process is based on Toomre's criterion for the radial instability of
gaseous discs (cf. \citealt{MDW04} for more
details). Since we used a log--polar grid, we have improved the
pre-computation of self-forces by subdividing each cell in $(n_r,
n_\phi, n_z)=(32,6,6)$ subcells. Self-forces are then linearly
interpolated before being subtracted from gravitational forces.

\begin{figure}
\resizebox{\hsize}{!}{\includegraphics{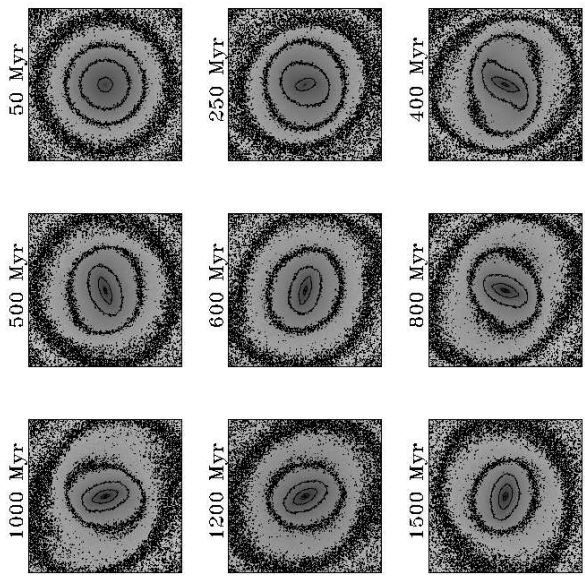}}
\caption{Evolution of the central $\pm 12$~kpc of \FBDS}
\label{fig:contour}
\end{figure}

When star formation is active, the radiative cooling of the gas has
been computed assuming a solar metallicity. We have also checked that
our results are independent of the radial resolution in the central
100~pc by using $N_R=40$ cells. In Fig.~\ref{fig:contour} we display
the region of the bar for various snapshots.  More details on the
global evolution of such kind of simulations, with similar initial
conditions and star formation on a longer timescale can be found in
\citet{MDW04} for instance.  

Compared to the simulations used in Paper~II, those described in the
present paper are much more accurate since we have used 5 times more
stellar particles and an improved N$-$body code. We will only deal
with a case extract from a dozen of such simulations of various
resolutions and initial setup. However, all our simulations display
\sdrops. The generic simulation \FBDS\ is thus representative of our
database. Another example of simulations with \sdrops\ are given in
\citealt{n1068} for a study of NGC\,1068.

Apart from the smaller number of stellar particles ($1.1\,10^6$
particles), the setup of the stellar and gas distribution of run
\EACB\ is similar to others runs. It includes an additional live dark
halo made of $2.2\,10^6$ live particles distributed in a Plummer sphere
of scalelength $50$~kpc and mass $6.46\,10^{11}$~M$_{\sun}$. As our
intention is to check the robustness of our results against the
presence of the massive dark halo, a Plummer sphere is a simple but
sufficient description of the halo. The setup of initial velocities
and velocity dispersions of disc particles is made consistently with
the dark matter distribution.  Since the interaction between the
extended live dark matter spheroid and the stellar/gaseous disc cannot
be computed with our particle--mesh code because of the limited
vertical range of the grid code, run \EACB\ has been evolved with {\tt
GADGET}, the tree-based N-body$+$SPH code developed by
\citet{gadget}. We used a adaptive time step based on the dynamical
time and limited by the Courant--Friedrichs--Levy condition.

\EACB\ does not exhibit any significant stellar \sdrop\ up to 5.35~Gyr
although others kinematical and morphological stellar properties are
roughly similar to halo-dominated models as those described by
\citet{AM02} (e.g. flat rotation curve, stellar ring encircling the
bar, etc.). The stellar radial velocity dispersion is indeed rather
flat, at $\approx 180$~\kms\ in the central 2~kpc at $t=5.35$~Gyr. On
the contrary, the radial velocity dispersion of the gas drops by
20~\kms\ in the central 8~kpc, approximately the diameter of the
stellar bar. Others runs, not displayed here, with more massive and/or
more extended halos do not develop stellar \sdrops\ either.  Since our
star formation recipes have not yet been implemented in {\tt GADGET}
we are not able to check the combined effect of a live dark halo and
star formation.  The effect of a halo on the \sdrop\ formation thus
deserves a much more in-depth study than the present one. This will be
discussed in a future paper.

\begin{table}
\flushleft
%\centering
%\begin{minipage}{126mm}
\caption{List of runs}
\label{tab:simul}
\begin{tabular}{@{}llllllll@{}}
\hline 
Model & Code & End & SF stop & $M_1$ & $M_2$ & $M_g$ \cr
      &      &(Myr)& (Myr)   &       &       &       \cr
\hline
\FBDT& PM    &2100 & no SF   & 0.055  & 0.55  & $\ldots$  \cr
\FBDS& PMSPH &2100 & on      & 0.05   & 0.5   & 0.055\cr
\FBDV& PMSPH&  643 & 400     &    "   &    "   &   "    \cr
\FBDW& PMSPH&  801 & 600     &   "    &   "    &    "   \cr
\FBDX& PMSPH& 1612 & 1000    &    "   &   "    &    "   \cr
\EACB&GADGET& 5350 & no SF   &    "   &   "    &   "    \cr
\hline
\end{tabular}
%\end{minipage}
\end{table}

%--------------------------------------------------------------------
% EVOLUTION DES DROPS DANS LES SIMULATIONS HYBRIDES
%--------------------------------------------------------------------
\section{Long term evolution of a \sdrop}
\label{sec:evolution}

To estimate the amplitude of a \sdrop\ (hereafter $\Delta\sigma$), we
have chosen to express it as the difference between two representative
or mean values of $\sigma$: one outside the drop in the kinematically
hot part of the bulge (\smax) and one inside the drop, i.e. the
central mean $\sigma$ (\smin). Moreover, we will only deal with
$\sigma_R$ for the sake of simplicity since the other two components
of the stellar velocity dispersion ($\sigma_\theta, \sigma_Z$) also
display the same effects (cf. Paper~II). Every value of $\sigma_R$ is
obviously a mass-weighted average since new particles may have
different mass than initial ones.

To determine the value of \smin, we have selected a central cylinder
of radius 100~pc (i.e. the central 200~pc) at each time. We made
several trials with height-limited disc instead of a cylinder but we
found no significant differences apart from a higher noise due to
lower numbers of particles. The selection typically leads to particle
numbers between 4000 and 10000 particles in the central 200~pc. 

After some trials, we decided to compute \smax\ in two annuli. The
first one, between 450 and 550~pc, represents a 'quiet' region,
outside the drop and the close surrounding region, but still in the
region that could be considered as the bulge of the galaxy. The second
one has been chosen between 250 and 350~pc as being representative of
the maximum value of $\sigma_R$ outside the central drop. The
difference between \smax\ in such a region and \smin\ give an extreme
value of \sdrop. However, the radius of the maximum $\sigma_R$
slightly increases during the evolution. We thus found more practical
to only use the first annulus (450$-$550~pc) to analyse into some
details the typical \sdrop\ evolution although it slightly
underestimates the real \smax\ and thus the real depth of the \sdrop.
%and only use the second annulus to illustrate the
%effect of switching off star formation.

\begin{figure}
\resizebox{\hsize}{!}{\includegraphics{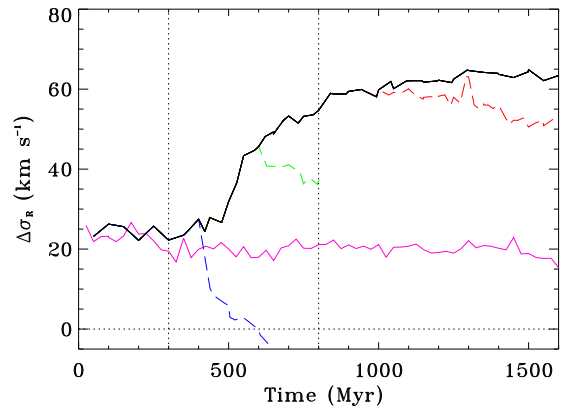}}
\caption{Evolution of the \sdrop\ amplitude estimated as the
difference between $\sigma_R$ averaged between 450 and 550~pc and the
mean $\sigma_R$ inside the central 200~pc. The reference run \FBDS\ is
plotted as a full black line. Star formation has been switched off for
runs \FBDV\ (blue), \FBDW\ (green) and \FBDX\ (red) at, respectively,
400, 600 and 1000~Myr (dashed lines). Vertical dotted lines show where
star formation has been switched off. Run \FBDT\ is also plotted as a
full magenta line for reference}
\label{fig:deltasigma}
\end{figure}

$\Delta \sigma$ is displayed in Fig.~\ref{fig:deltasigma}. Its
evolution could be roughly separated into three phases. Boundaries of
these phases are approximately defined since they should depend on the
details of the star formation history in the central region.

\subsection{$0-300$~Myr} 
$\Delta\sigma$ remains approximately constant at 25~\kms. The
comparison with the pure N-body run \FBDT\ shows that such an
amplitude is expected in case of a low star formation rate or no star
formation at all. Indeed, it has been shown in Paper~II that 
% HW 2/3/2006
%{\bf
an artificial \sdrop\ of such a small amplitude is expected because of
N-body softening effects. However, the amplitude of such a numerical
\sdrop\ is much lower than those observed and should fully diseappear
after convolution by any realistic point spread function. 
%}

\begin{figure}
\resizebox{\hsize}{!}{\includegraphics{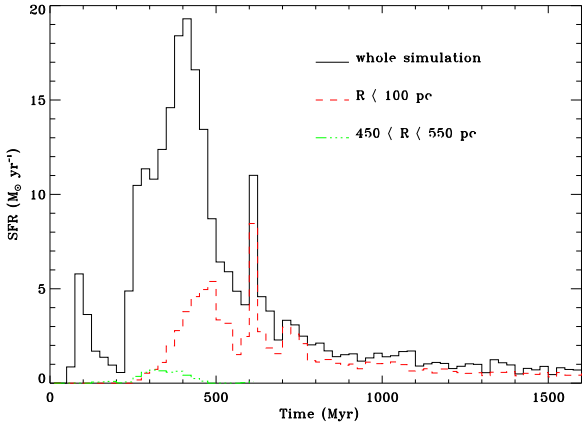}}
\caption{Evolution of the star formation rate for run \FBDS. SFR is
displayed for the whole simulation (full line), the central 200~pc
(dashed line) and the annulus of radii 450-550~pc (dashed-dotted
line)}
\label{fig:sfr}
\end{figure}

\begin{figure}
\resizebox{\hsize}{!}{\includegraphics{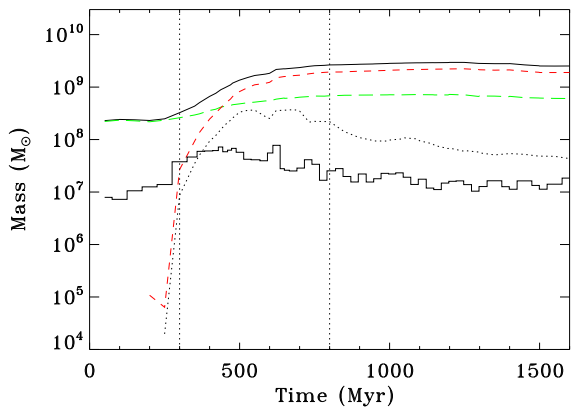}}
\caption{Evolution of the mass inside the central 200 pc for run
\FBDS. The black line is the total mass, the green long-dashed line is
the mass of the old population, the red short-dashed line is the mass
of the new population, the histogram-like curve is the mass of the
gas. The black dotted line is the mass of the new population that is
born inside the central 200 pc and is still in this region for the
time given in abscissa. Vertical dotted lines show the three phases
discussed in the text }
\label{fig:mass}
\end{figure}

The SFR inside the central 200~pc is displayed in
Fig.~\ref{fig:sfr}. For $t < 300$~Myr the central SFR remains below
1~\sfrunit while most of the new particles are created in the large
scale disc. Thus, at $t=300$~Myr, the mass of the new population
roughly amounts to that of the gaseous component but remains an order
of magnitude lower than the old population
(cf. Fig.~\ref{fig:mass}). 
% calcul: sfr_fbds, sfr_spec
% figure: figsfr
%It is
%noteworthy that the local SFR maximum in the central 200~pc ($\approx
%5$~\sfrunit\ at $t\approx 475$~Myr) is delayed with respect to the
%total SFR maximum by $\approx 75$~Myr (cf.~Fig.~\ref{fig:sfr}). 

% calcul: evol_mass
% figure: figmass

\subsection{$300-800$~Myr} 
The \sdrop\ increase is now very significant so that $\Delta\sigma$
reaches $\approx 55$~\kms. This is the \sdrop\ growth phase which
corresponds to the burst of star formation in the central
region. However, such phase could be due either to the decrease of
\smin\ or to an increase of \smax.

\begin{figure}
\resizebox{\hsize}{!}{\includegraphics{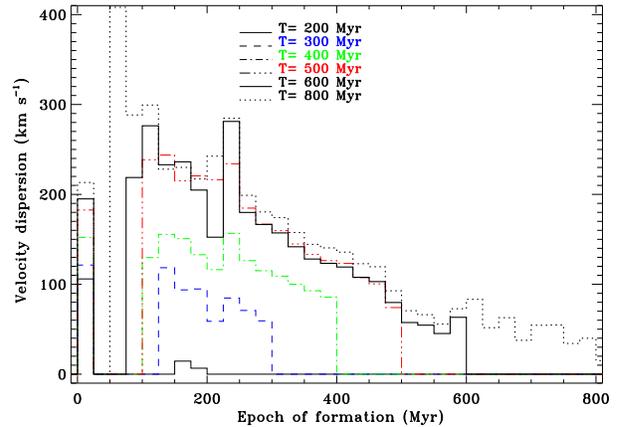}}
\resizebox{\hsize}{!}{\includegraphics{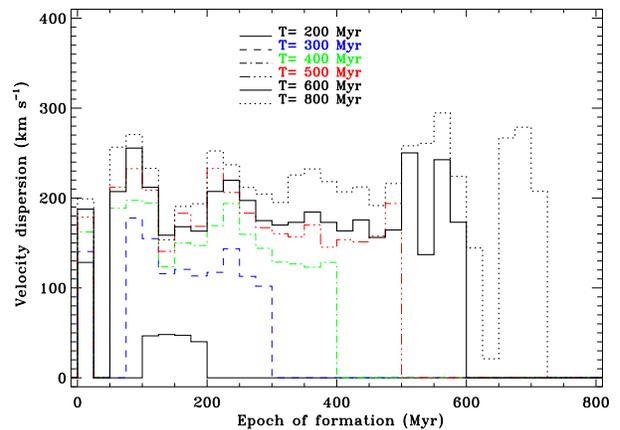}}
\caption{Distribution of $\sigma_R$ as a function of the epoch of
formation for particles selected within a radial range [0, 100] pc
(top panel) and [450, 550]~pc (bottom panel). Bin size is 25~Myr}
\label{fig:sigmaage}
\end{figure}

To disentangle both effects, we display in Fig.~\ref{fig:sigmaage} the
distribution of $\sigma_R$ as a function of the epoch of particle
formation for several selected times, up to $t=800$~Myr. The two
regions of reference are separately displayed. The old population is
included in the first bin (0-25~Myr). For each curve, the last bin
represents the mean $\sigma_R$ of the newest particle born during
25~Myr before the corresponding time.

The $\sigma_R$ distribution in the central 200~pc clearly shows an
increase with time between $t=200$ and $t=600$~Myr, for all generation
of particles. The newest particles have a lower $\sigma_R$ than the
old population as expected. At each time, the shape of the histogram
is roughly a linear relationship between $\sigma_R$ and the epoch of
particle formation. In first approximation, the rate of heating is
independent of the epoch of formation, i.e. the age of particles. This
heating process stops around $t=600$~Myr. The $\sigma_R$ distribution
then does not evolve any more for particles born between $t=100$ and
$t=600$~Myr. A number of particles born for $50< t< 100$~Myr gets into
the central region from the outside; these particles have a very high
$\sigma_R$, greater than the old population, but their total mass is
not high enough to increase \smin.

\begin{figure}
\resizebox{\hsize}{!}{\includegraphics{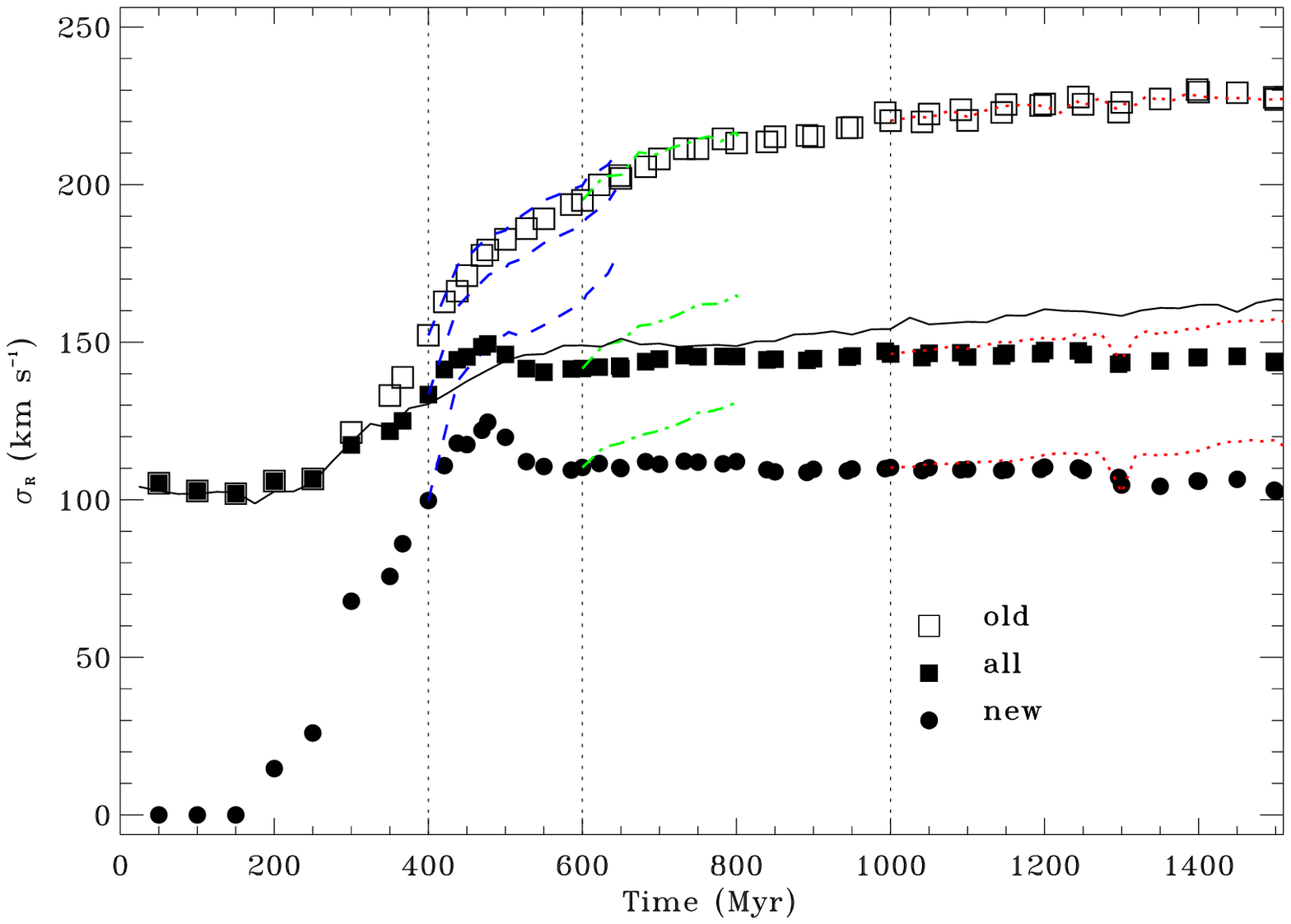}}
\resizebox{\hsize}{!}{\includegraphics{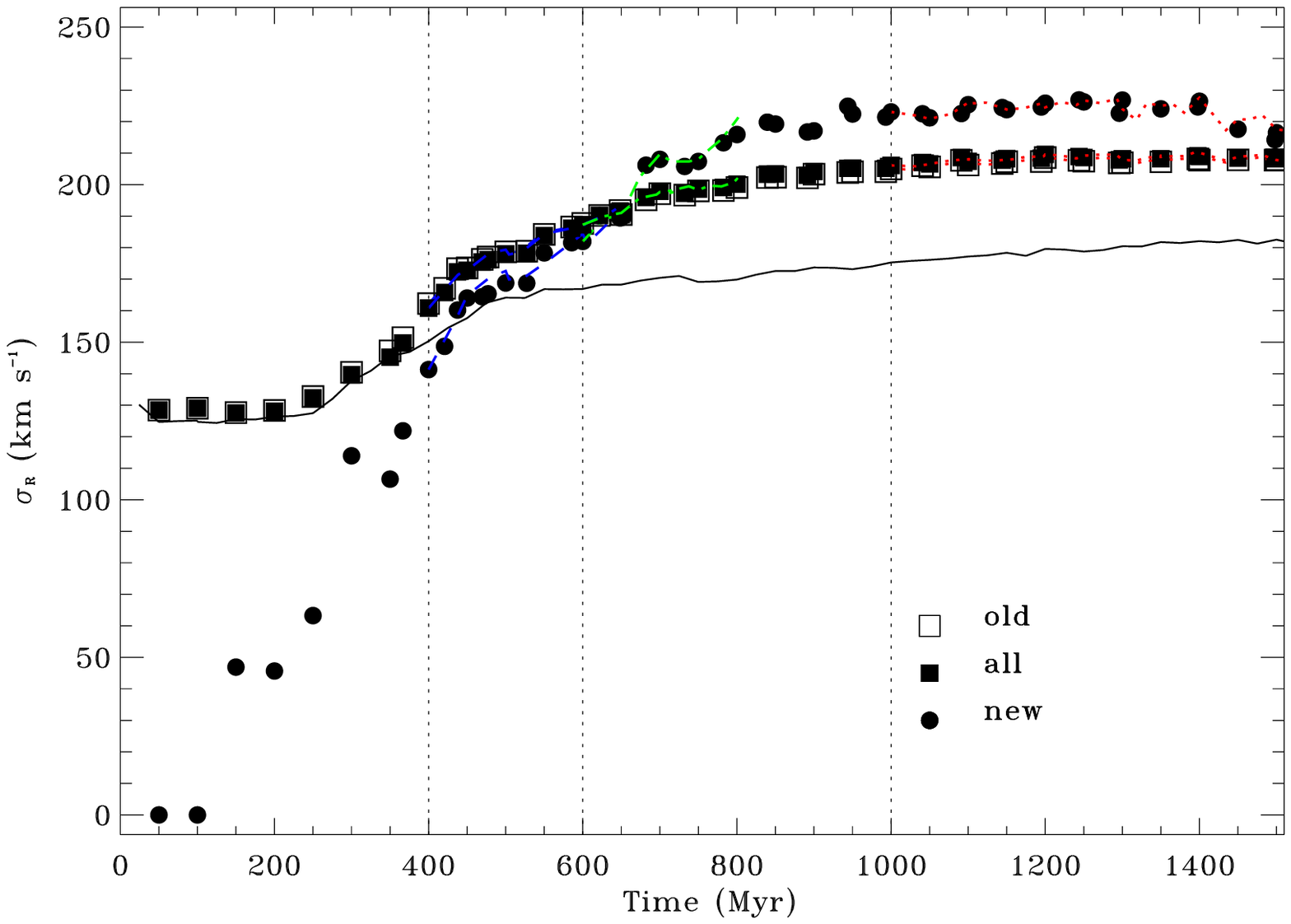}}
\caption{Evolution of $\sigma_R$ for particles selected within a
radial range [0, 100] pc (top panel) and [450, 550]~pc (bottom panel).
Symbols represent \FBDS. Opened squares are the velocity dispersion of
the initial population (\sold), full squares are the total $\sigma_R$
and full circles are the velocity dispersion of the new particles
(\snew) created by star formation during the simulation.  Runs \FBDV\
(blue dashed lines), \FBDW\ (green dotted-dashed lines) and \FBDX\
(red dotted lines) can only differ from \FBDS\ when star formation is
switched off after respectively $t=400$, $t=600$ and $t=1000$~Myr
(times are marked by dotted vertical lines).  $\sigma_R$ for the pure
N$-$body run \FBDT\ is plotted as a full line}
\label{fig:sfoff}
\end{figure}

Another way to represent \smin\ evolution is to display separately the
evolution of $\sigma_R$ for the two particle populations of run \FBDS\
(\sold\ for the initial or `old' particles and \snew\ for the `new'
particles, i.e. for particles born during the run) and for all the
particles (Fig.~\ref{fig:sfoff}). The control run \FBDT\ is also
plotted to show the heating of collisionless particles for $t\ga
250$~Myr. Such heating is mainly due the development of the stellar
bar perturbation which increases the mass in the central region
(cf. Paper~II). Run \FBDT\ cannot develop a \sdrop. However, this run
allows us to estimate the increase of $\sigma_R$ due to the secular
heating of the central region.

Fig.~\ref{fig:sfoff} (top panel) shows that \smin\ clearly increases
until $t\approx 475$~Myr and then slightly decreases to reach a
plateau around \smin$\approx 140$~\kms. The stellar mass in the
central 200~pc is mainly made out of new particles, especially for $t
\ga 400$~Myr (Fig~\ref{fig:mass}). \smin\ variations are thus strongly
correlated to \snew\ ones.  The population of new particles
responsible for this plateau (those that are born at $600 < t <
800$~Myr) is made out in part of particles born inside the central
200~pc and in part of low $\sigma_R$ particles that have migrated from
the outside. The fraction of new particles which is born and stays in
the central region vary from 25 to 10\%\ between $t=600$ and
$t=800$~Myr. Thus, the migration of low $\sigma_R$ particles from
outside the central 200~pc takes over from particles born inside the
central region when the local SFR strongly decreases after $t=475$~Myr
(cf. Fig.~\ref{fig:sfr}). This dynamical effect is obviously due to the
strong growth of the central total mass which sinks the potential
well.

The evolution of \smax\ is not driven by the same causes. Indeed, the
$\sigma_R$ distribution in the 450-550~pc annulus
(Fig.~\ref{fig:sigmaage}, bottom panel) does not show any saturation
of the heating process. On the contrary to the central region, there
is no clear relationship between $\sigma_R$ and the epoch of
formation. This is obviously due to the fact that new particles that
are born in the annulus region cannot stay there.
%Moreover, the gas
%$\sigma_R$ is higher in the annulus than in the central region so that
%the newest particles born with typical $\sigma_R$ higher than in the
%central region. 
Thus, both \sold\ and \snew\ increase. As a consequence, \smax\
continuously grows (Fig.~\ref{fig:sfoff}, bottom panel), its values
being moreover greatly dominated by \sold\ since the old population
represents most of the mass in the annulus.

The \sdrop\ growth is thus due in great part to the heating of the
surrounding region, since \smin\ reaches a plateau.  

\subsection{$ > 800$~Myr}
\begin{figure}
\resizebox{\hsize}{!}{\includegraphics{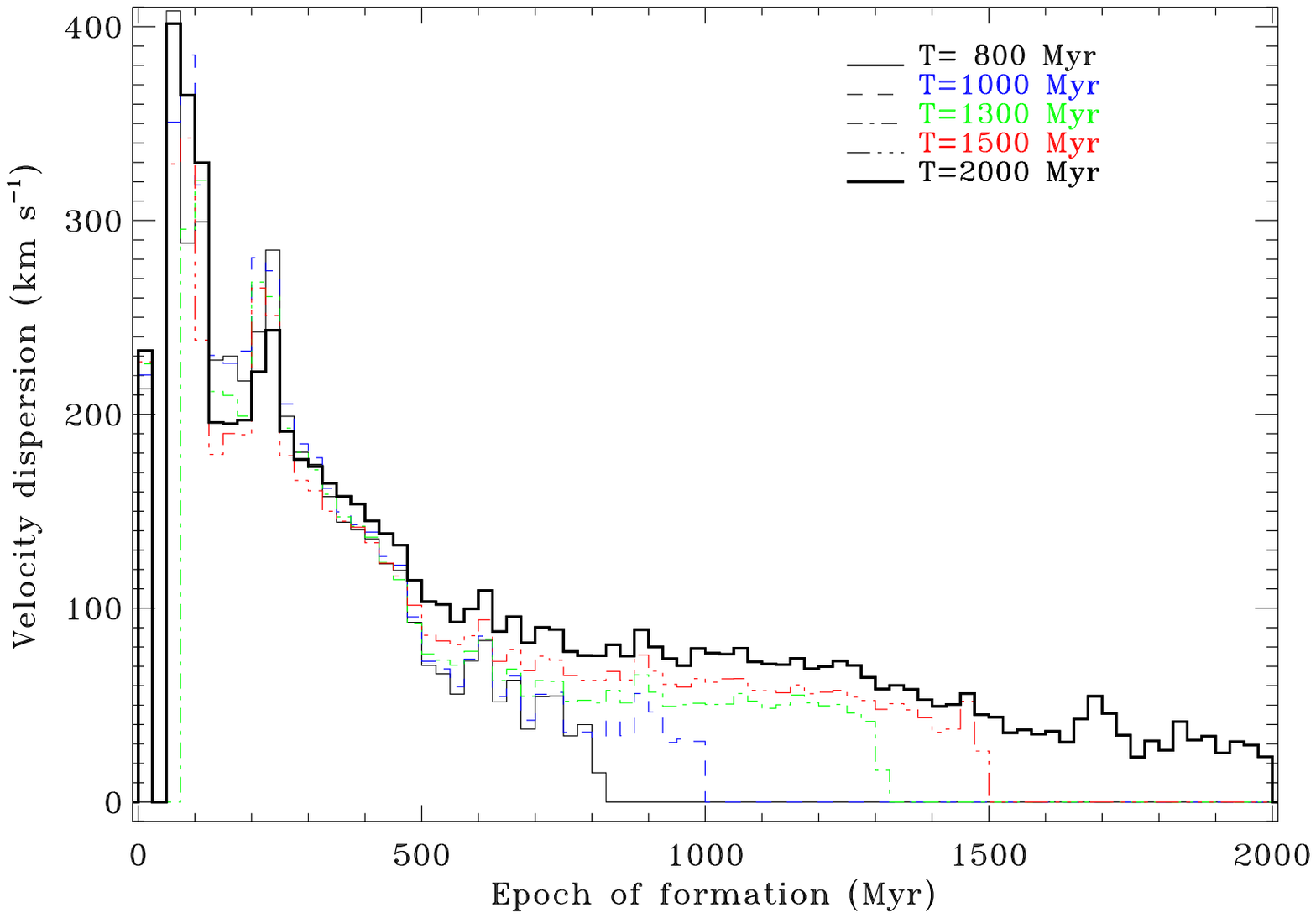}}
\resizebox{\hsize}{!}{\includegraphics{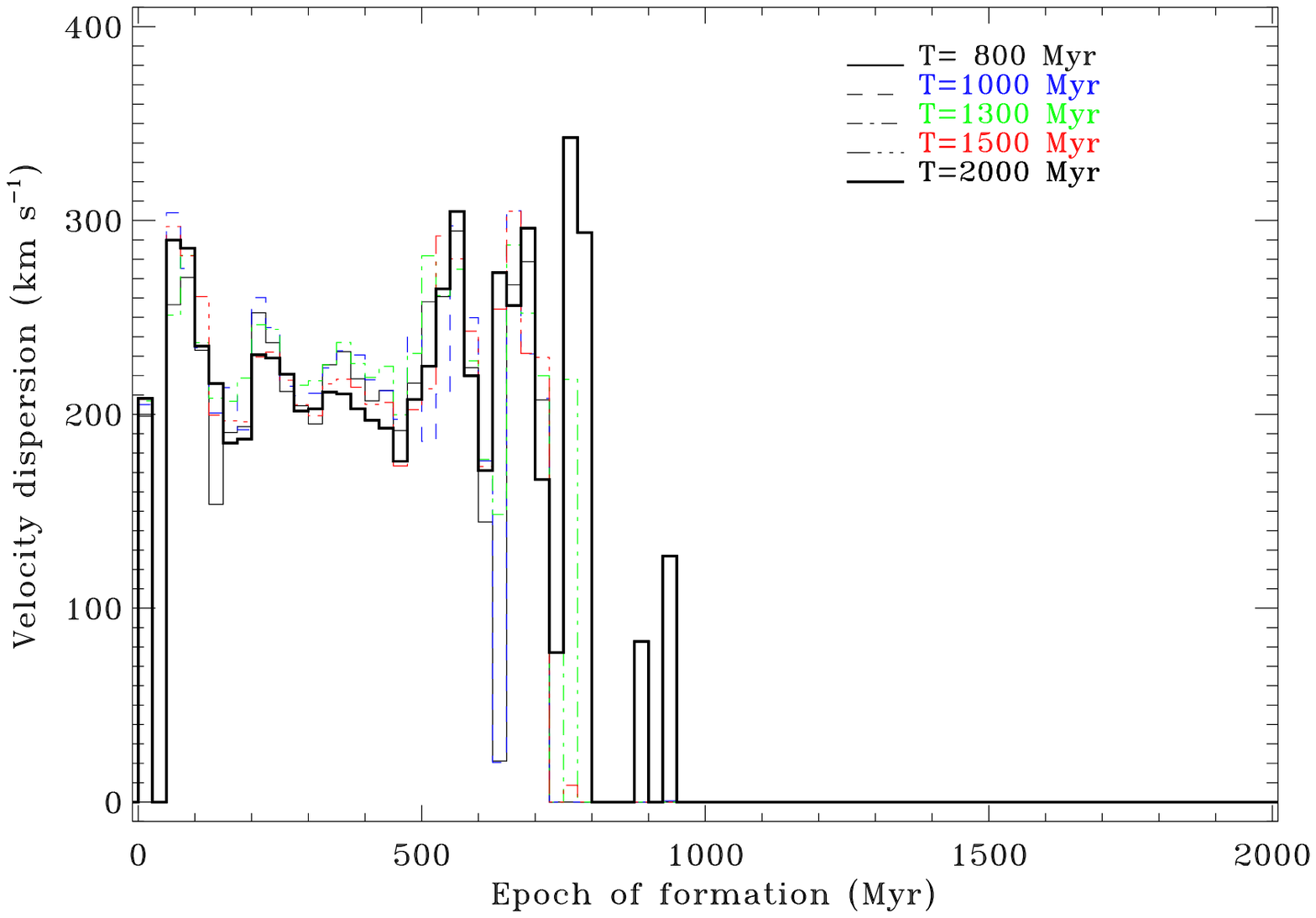}}
\caption{As Fig.~\ref{fig:sigmaage} but for $t \ge 800$~Myr. The
$t=800$~Myr curve is the same than for Fig.~\ref{fig:sigmaage}}
\label{fig:sigmaageb}
\end{figure}

During this phase, the \sdrop\ development seems to saturate. Indeed,
$\Delta\sigma$ gains only 10~\kms\ in roughly 700~Myr. In the central
200~pc, after $t=800$~Myr, the slope of the relationship between
$\sigma_R$ and the epoch of formation changes for particles born after
$t\approx 500$~Myr (cf. Fig~\ref{fig:sigmaageb}). During this phase,
the heating is less efficient so that particles born after $t\approx
500$~Myr keep a low $\sigma_R$. Moreover, most of the mass in the
central region is due to particles born between 250 and 600~Myr,
i.e. during the burst of star formation (Fig.~\ref{fig:sfr}). Thus,
\smin\ does not evolve (cf. Fig~\ref{fig:sfoff}), as \snew, even if
\sold\ continues to slightly increase.

The 450-550~pc annulus being dominated by the old population, \smax\
is driven by the behaviour of \sold\ which slightly increase by a few
\kms. It is noteworthy that \snew\ is greater than \sold\ during this
phase. This effect is mainly due to particles born between 50 and
625~Myr which cross this region.  Finally, since \smin\ does not
evolve, the small \sdrop\ fluctuations are again mainly due to the
weak heating of the surrounding region.

For $t > 1300$~Myr, $\Delta \sigma$ fluctuates around a mean value of
65~\kms, with no significant sign of increase or decrease. Thus, a
\sdrop\ is not a transient feature.

\section{Effect of switching star formation off}
\label{sec:sfoff}

To throw some light on the matter of \sdrop\ lifetime, we have
switched off the star formation process at three different times of
run \FBDS\ and pursued the runs (runs are named \FBDV, \FBDW\ and
\FBDX) for a while. For \FBDV, star formation has been stopped at
$t=400$~Myr, when the total star formation rate (SFR) is maximum
(roughly 20~\sfrunit, cf. Fig.~\ref{fig:sfr}). The simulation was then
proceeded until $t\approx 650$~Myr. For \FBDW, star formation has been
stopped in the middle of the \sdrop\ growing phase ($t=600$~Myr)
whereas it has been switched off at $t=1000$~Myr for \FBDX, when the
\sdrop\ is well developed and closed to be stable.

The first obvious effect of switching off the star formation is the
decrease of $\Delta\sigma$ for all the three runs
(cf. Fig.~\ref{fig:deltasigma}). For run \FBDV, $\Delta\sigma$ becomes
even negative for $t > 600$~Myr which means that the central region
becomes hotter than the rest of the galaxy. The heating timescale is
very short.

As in Sect.~\ref{sec:evolution}, we need to disentangle $\sigma_R$
variations for both representative regions and thus come back to
Fig.~\ref{fig:sfoff} where \smin\ and \smax\ of \FBDV, \FBDW\ and
\FBDX\ are compared to those of \FBDS, as well as the contribution of
\sold\ and \snew\ to \smin\ and \smax.

For $t > 400$~Myr, the mass in the central 200~pc is dominated by the
new population when star formation is active, so that \smin\
closely follows \snew. Indeed, the mass of the new population
continuously increases when star formation is active
(Fig.~\ref{fig:mass}) so as to equals the mass of the old population
at $t\approx 440$~Myr. Then, the new population dominates the central
region. Obviously, this is no more the case when star formation is
switched off. Indeed, between $t=400$ and $t=600$~Myr, \smin\ of
\FBDV\ increases by $\approx 55$~\kms. The rate at which \snew\
increases is higher ($\approx 0.8$~\kms\ Myr$^{-1}$) during the first
$\approx 50$~Myr than the following 150~Myr ($\approx 0.1$~\kms\
Myr$^{-1}$, very close to the rate of run \FBDT).  A marginal effect
is the small heating of the old population ($\approx 10$~\kms\ at
$t=600$~Myr). The heating of the new population is also very efficient
in the 450-550~pc annulus but, since this region is dominated by the
old population, the effect on \smax\ is limited to $\approx
25$~\kms. Thus the heating of the central region is mainly responsible
to the \sdrop\ disappearance.  This heating has mainly two causes:
\begin{enumerate}
\item the rapid gas accumulation in the central region strongly
increases the central mass and makes local perturbations of the
gravitational potential;
\item low-$\sigma$ particles that escape from the central region are
not replaced by new ones. Indeed, the new population is made only in
part of particles which are born and remain inside the 200~pc.  We show in
Fig.~\ref{fig:mass} how evolves the mass of this sub-population. It
clearly decreases which means that a fraction of low-$\sigma$
particles could be heated outside the central region, during their
escape. When star formation is active, they are replaced at each
timestep by the last generation of particles. This is no more the case
for \FBDV, \FBDW\ and \FBDX.
\end{enumerate}

For \FBDW\ and \FBDX, the decrease of $\Delta\sigma$ has a longer
timescale than for \FBDV\ even though the same mechanism than for
\FBDV\ comes into play when the star formation is switched off. The
amplitude of the decrease of $\Delta\sigma$ is therefore smaller for
\FBDW\ than for run \FBDV, and even smaller for \FBDX. Since the
central \snew\ is roughly constant for $600 < t < 1000$~Myr for \FBDS\
(cf. Fig.~\ref{fig:sfoff}), one cannot invoke the effect of a colder
population which needs more energy to be heated. In fact, the heating
of the central region is much more difficult at $t=600$ or
$t=1000$~Myr essentially because the mass of the new low-$\sigma$
population is greater than the old one (Fig.~\ref{fig:mass}). Another
cause is the bar itself. For $t > 475$~Myr, the gravitational
perturbations made by the formation of the stellar bar have
vanished. Finally, there is also less gas accumulation in the centre
than for \FBDV\ since star formation has consumed a significant
fraction of the available reservoir in the central region.

%--------------------------------------------------------------------
% DISCUSSION
%--------------------------------------------------------------------
\section{Conclusions}
\label{sec:conclusions}

We confirm the result found in Paper~II that the appearance of a
stellar \sdrop\ in hydro$+$N$-$body simulations is unavoidable on the
condition that star formation is take into account. Indeed,
low-$\sigma$ gas accumulates into a central disc because of the
gravitational torques induced by the stellar bar. It gives raise to a
new stellar population with the same velocity dispersion. Therefore,
the \sdrop\ appears quickly after the beginning of the bar formation,
i.e. $\approx 300$~Myr for our simulation \FBDS. It takes roughly
1~Gyr to reach the maximum amplitude. Then, the amplitude of the
\sdrop\ remains constant ($\ga 65$~\kms) until the end of the
simulation (2.1~Gyr). Timescales could however strongly depend on the
star formation rate in the central region.

A \sdrop\ is thus not a transient feature. However, all stellar
populations, even low-$\sigma$ particles, are heated, probably by
fluctuations in the central region of the gravitational potential
field as those described by, for instance, \citep{CS85}, or stochastic
heating made by resonances between several density waves \citep{MQ05}.
However, in this paper, we did not investigate into details what could
be the heating mechanisms since this deserves a dedicated study.  The
\sdrop\ persistence is thus not due to some improbable cooling
mechanism.  Our experiments of switching off the star formation
recipes at various epoch prove the relationship between this long
lifetime and the refurbishment of the new stellar population of
low-$\sigma$. A star formation rate as low as 1~\sfrunit\ in the
central 200 pc is enough to sustain a permanent \sdrop\ by replacing
heated particles by new low-$\sigma$ ones. Such a rate is not uncommon
in barred galaxies provided that the central region is regularly fed
by fresh gas.  Moreover, when the star formation rate is too low, it
will take several hundred Myr for the \sdrop\ to disappear. This could
explain why \sdrops\ are still observable in early-type galaxies with
no sign of nuclear star formation (e.g. NGC\,4477).

The visibility of a \sdrop\, as well as its opposition to heating
mechanisms, strongly depends on the mass ratio between the old `hot'
population and the `new' cold one. For \FBDS\ the central star
formation rate reaches 5~\sfrunit\ during the bar formation phase so
as to create a new population which becomes more massive than the
initial one at $t \approx 440$~Myr, thus on a rather short
timescale. \sdrops\ should thus be rather common in barred galaxies in
the local universe. A systematic survey of the central kinematics in
barred galaxies could confirm these results.

%--------------------------------------------------------------------
% ACKNOWLEDGEMENTS
%--------------------------------------------------------------------
%\begin{acknowledgements}
\section{acknowledgements}
We are grateful to E. Emsellem and P. Ferruit for fruitful
discussions.  Our computations were partly performed on the IBM-SP4
hosted by IDRIS/CNRS and the CRAL 18 nodes cluster of PC funded by the
INSU/CNRS (ATIP \# 2JE014 and Programme National Galaxie).
%\end{acknowledgements}

\end{document}